# Deep learning based mixed-dimensional GMM for characterizing variability in CryoEM

Muyuan Chen[1] and Steven J. Ludtke[1]

[1] Verna Marrs and McLean Department of Biochemistry and Molecular Biology, Baylor College of Medicine, Houston, Texas, USA

**Abstract**

Structural flexibility and/or dynamic interactions with other molecules is a critical aspect of protein function. CryoEM provides direct visualization of individual macromolecules sampling different conformational and compositional states. While numerous methods are available for computational classification of discrete states, characterization of continuous conformational changes or large numbers of discrete state without human supervision remains challenging. Here we present e2gmm, a machine learning algorithm to determine a conformational landscape for proteins or complexes using a 3-D Gaussian mixture model mapped onto 2-D particle images in known orientations. Using a deep neural network architecture, e2gmm can automatically resolve the structural heterogeneity within the protein complex and map particles onto a small latent space describing conformational and compositional changes. This system presents a more intuitive and flexible representation than other manifold methods currently in use. We demonstrate this method on both simulated data as well as three biological systems, to explore compositional and conformational changes at a range of scales. The software is distributed as part of EMAN2.

**Introduction**

Cryo-electron microscopy (CryoEM) is used to image biological macromolecules in a near-native state and is ostensibly capable of resolving structures to near-atomic resolution. However, most macromolecules possess substantial conformational and/or compositional variability as part of their biological function. In single particle reconstruction (SPR), a micrograph contains a snapshot of many macromolecules, each frozen at a random point on its conformational and/or compositional landscape. This presents the difficulty that the features visible in any single structure solved using CryoEM data will be limited by the conformational variability among the particles making it up. With more complete analysis, the presence of these variations can be turned into an advantage, as the individual data intrinsically explore a large portion of the conformational landscape of the system. With appropriate methods, achieving an ensemble of individually more detailed structures could be achieved.

Many methods have been developed to address the heterogeneity problem in SPR[1–3]. Perhaps the oldest and most commonly used method is multi-model refinement/3-D classification, in which multiple 3-D maps are used as references and each particle is compared to the projections of each reference in each iteration[4–7]. Focused classification is a variant of this method in which variability is explored only inside a user-defined mask[8]. These methods often work quite well when the system falls into a small number of discrete states, such as the two states associated with ligand binding. However, to work well, the number of discrete states should be small, and the quality of the initial seed volumes often has an impact on the results.

Another common practice is to perform multi-model refinement, then rely on a human to discard particles representing states judged to be "bad"[9,10]. This process is typically iterated multiple times until a single map with improved resolution is achieved. This method ostensibly produces one structure at high resolution representing the most populous state in the data, by simply ignoring any contradictory data. This has the disadvantage of imposing potential human bias on the results, and while the resulting map generally has improved resolution, it clearly presents an incomplete picture of the macromolecule being studied.

In addition to 3-D classification, multi-body refinement can be used to resolve local structural variability caused by conformational changes[11]. This technique relies heavily on researchers' prior knowledge of structural domains in protein and requires the regions of interest to be large enough to provide sufficient signal for local orientation assignment.

Finally, we have seen the recent emergence of manifold embedding techniques to address the problem of structural variability[12–16]. These methods are fairly new and varied in their mathematical methods. While they have shown promising results, they also face difficulties in mapping the manifolds to biological interpretation, and the process of interpreting the structure at a point on the manifold is often time consuming.

In this manuscript we present a strategy leveraging deep learning technology to map 2-D data directly to a 3-D Gaussian mixture model (GMM). This produces a representation where conformational and compositional variations can be directly and intuitively related back to the data representation. A point on the manifold represents a specific Gaussian configuration which can be instantaneously visualized, without needing to first reconstruct large subsets of particle data.

**Methods**

One of the difficulties in SPR heterogeneity analysis is the mathematical representation of protein conformations. If we consider the motion of an object from position A to position B along a simple linear path, it should be possible to represent the position of the object on the path with a single value. However, when we represent this motion via images or volumes, the motion becomes a pattern of pixels becoming brighter and dimmer along the path in a complex sequence. Simple image analysis methods, such as principal component analysis (PCA), can readily identify the pixels involved in such a motion, but cannot readily map the highly nonlinear sequence of pixel variations back to the single degree of freedom we know exists in the underlying system.

The second difficulty lies in the fact that in single particle analysis, variations occur in 3-D space, and yet the individual measurements are 2-D projections, lacking information about the motion in the 3rd dimension. However, a sufficiently large ensemble of images in different orientations and states typically contains sufficient information to constrain both the 3-D structure and the motion.

Rather than attempting to determine paths in image space, we instead impose a Gaussian model at a specified level of detail, and then identify changes in Gaussian location and intensity which are self-consistent with the ensemble of particles. In this Gaussian mixture model (GMM) each function is defined by five variables: 3-D coordinates, amplitude, and width. The width parameter is typically fixed, representing the level of detail in the representation. In this approach, the local motion of a domain would be represented by a simple change in location of the Gaussians making it up, and the presence/absence of a ligand would be represented by a change of amplitude.

Converting the Gaussian representation into an image representation (a projection) is a trivial process, whereas the inverse process, mapping a single image to a set of 3-D Gaussian locations, is underdetermined. The inverse problem is sufficiently constrained only when a large image ensemble is considered. To solve this sparse and nonlinear inverse problem, we make use of deep learning methodologies (Fig 1). This design requires only the definition of a loss function describing the agreement between individual images and specific configuration of 3-D Gaussians. We make use of the Fourier ring correlation (FRC) metric[17] in the loss function, which has the additional advantage of being insensitive to microscope contrast transfer function (CTF) artifacts so long as the image is reasonably stigmated with minimal drift, and phase-flipping corrections have been applied to the particle images.

The network design involves 2 components. First, a decoder, which maps a small latent vector, into a set of 5N Gaussian parameters, where N is the number of Gaussians. The latent vector is simply a reduced dimensionality representation of the 3-D configuration of the molecule. In linear analysis, each component in the latent vector can represent one degree of freedom in the macromolecule. However, with the nonlinearity provided by the network, it is possible for local regions in the latent space to represent independent discrete states.

The second network component is the encoder, which maps 2-D images, via their derivatives, into latent vectors. The latent vector then passes through the decoder to produce 5N Gaussian parameters, which immediately provides a 2-D projection or 3-D volume as desired. This mapping process is constrained by the latent vector representation, and the set of particles mapped into this latent space will form a manifold, conceptually similar to other manifold methods in CryoEM. However, due to the nonlinearities and our enforcement of a GMM with a specified level of detail, it also becomes possible to probe systems in very specific ways, which would be difficult using competing methods. For example, parameters of specific Gaussian components can be held constant, such that the GMM considers only variability in specific regions, or looks for correlations between specific regions.

Our network structure is conceptually similar to an autoencoder[18], in which the network is trained directly from raw data, with no need for ground truth. The goal in an autoencoder is for the network optimally match the input data to itself at the output, passing through a low dimensionality latent space in the center. In our case, the input data is 2-D particle data, and our network output is the full 3-D GMM. While this 3-D model can recreate 2-D projections for training, the GMM output is far richer than any individual 2-D image. To achieve this result, a slightly different network training strategy is required.

We begin by training only the decoder, such that it produces a single neutral 3-D structure from a latent vector with all components set to zero. The network is trained to produce 5N Gaussian parameters best matching the provided neutral structure. To avoid becoming trapped in a local minimum during training, we begin with a downsampled version of the map, then progressively increase the sampling as the training process converges. The decoder is trained using an ADAM optimizer[19] with the FRC between the GMM and the provided map as the loss function, with optional regularizers to promote uniform distribution. When trained, the decoder produces an accurate representation of the neutral map when given an input latent vector of zero.

Next, the encoder, which produces latent vectors from particle data, must be included in the training process. The goal of this procedure is for the latent space vector to represent as much of the variability of the specimen as possible. The training data consists of 2-D particles and their orientation parameters from a standard single particle refinement. The assigned orientations for the particles can be imported from a standard EMAN2 or Relion refinement[20,21]. For each particle, we compute the gradient of the loss function between the particle image and GMM with respect to the GMM parameters from the neutral model. These gradients, 5N parameters per particle, become the input to the encoder for that particle. The gradient vectors are computed in the coordinate system of the GMM, so they are intrinsically invariant with respect to translation and rotation of the raw particles. The loss function is, again, the FRC between the particle and Gaussian projection. For training, the encoder weights are initialized with small random values producing latent vectors close to, but not exactly, zero.

The particle data and Gaussian parameters clearly will not agree perfectly, due to both noise present in the 2-D particle images as well as the conformational and compositional variability in the specimen. As noise will be completely random within each particle, whereas the conformational and compositional variability will follow patterns represented in many particles, the latent space should preferentially train for variations actually present in the data. We do not require the orientations to be truly optimal at this point, as when one part of the structure is moving with respect to another, the concept of a single correct orientation does not exist. Once the complete network has been trained to represent the variations in the data with the given orientations, another training cycle can be run where the particle orientations are refined against the dynamic GMM (Fig S1). This process can be iterated, though in practice, it is unlikely to take more than one or two iterations before the orientations and variability parameters agree to the best extent possible.

With a traditional PCA representation of variability in image space[22,23], the dimensionality of even a simple motion within a structure will be high since the motion involves many pixels undergoing nonlinear variations in intensity. As discussed above, with the GMM representation each independent motion should require, at most, a single variable in the latent space. Thus, our default of a 4-D latent vector can represent at least 4 independent variations. However, given the nonlinearity of the system and the fact that molecular variation tends to be highly constrained, it is readily possible for a single variable to possess multiple features across its domain. Thus, once all of the particles have been mapped into the latent space via the encoder, it is necessary to perform analysis on the particle distribution within the latent vector space. Any dimensional reduction algorithm can be used as part of this analysis to facilitate visualization. PCA applied to the latent space is one straightforward approach for visualization and segmentation. Even with the nonlinearity of the network, we still have the constraint that similar configurations will lie close to each other in the latent space and less similar configurations will be further apart. That is, we still expect continuous variations in structure to appear on manifolds in the latent space. Any latent vector can be easily visualized either immediately as its GMM representation or by reconstruction of the particles falling in a local region in the latent space.

## Results

Here we consider three data sets which are publicly available through EMPIAR[24], each of which exhibits different types of variability. The majority of the observed variations are well known in each case, providing a validation of the method. We also observe some additional motions not reported in the original study, but generally consistent with our understanding of the underlying systems. As these are public data sets, experimental validation of these observations is clearly beyond the scope of this manuscript. Nonetheless, we believe these examples present the power of the method. Tests using simulated data are included in supplementary data as a demonstration of how the method can represent multiple motions efficiently and accurately (Fig S5).

**50S Ribosome assembly**

For this test, we used a L17-Depleted 50S Ribosomal Intermediate dataset (EMPIAR-10076)[25], to demonstrate the method's ability to identify discrete variability, such as partial complex formation/ligand binding. This data was also used in two other recent manifold method manuscripts, permitting qualitative comparison of results[14,15]. We began with a structure determined using normal single particle methods in EMAN2 to 3.3 Å using the entire dataset with the exception of obvious ice contamination (124,900 particles). This structure was low-pass filtered to 8 Å, then used to generate a GMM with 3082 Gaussians. The specific number of Gaussians was empirically determined, based on the targeted level of detail, and has little qualitative impact. Any Gaussians falling outside a specified mask can be excluded from the final model. Since most of the structural variations within this dataset are the presence/absence of individual ribosomal components, we initially permitted only the Gaussian amplitudes to vary. After training, we took the 4-D latent space vectors (Fig S3) and used UMAP[26] to reduce the space to 2-D in order to visually explore the structural variability of the system. From the embedded space, particles were clearly separated into six visible clusters. The particles in each of these clusters were then used to produce a 3-D reconstruction representing the cluster. The observed structural differences recapitulate known states[25] of ribosome assembly as shown in Fig 2.

While the points form clear clusters in the 2D conformation space, such classification only represents large scale structural differences, and more subtle compositional changes can also be observed from particles within the same class. For example, we manually selected three points along a line in one of the clusters with the central protuberance domain and reconstructed an averaged structure from the 2000 particles closest to each of the points in the embedded space. The resulting structures show the introduction of h68-70 and h76-78 of the 23s rRNA[25]. Interestingly, selecting 3 points along a nearly parallel line in a different cluster, one without the central protuberance domain, we observe the introduction of the same rRNA helices in the structures along the selected line (Fig 3b).

Finally, we examined conformational changes within the system. One of the factors that limits the resolvability of the averaged ribosome is the smearing effect of the dynamic central protuberance domain. To study this, we continued training the network with the Gaussian positions also permitted to vary in this domain, including only particles where the central protuberance domain is present. Note that the network model always includes the full set of 5 parameters for each Gaussian, but any of these components can be held constant. This additional analysis identified a clear tilting motion of ~8 degrees of this domain (Fig 3d).

**Spliceosome**

To test the performance of e2gmm on large scale conformational changes we made use of pre-catalytic spliceosome data (EMPIAR-10180)[27]. We began with the particle orientation assignments and averaged structure determined using EMAN2 to 4.6 Å using the full dataset (327,490 particles). As resolution in CryoEM is a measure of self-consistency rather than visible detail, it is possible to achieve relatively high measured resolutions even in the presence of significant motion blurring, even when the structure clearly lacks high resolution detail. The density map was lowpass filtered to 13 Å and represented by 2048 Gaussians. All Gaussian parameters were allowed to change in this example. We used PCA to reduce the neural network latent space to 2D after training for visualization of the subspace with the most significant variation. Compared to nonlinear dimension reduction methods, PCA conveniently preserves the inverse transform, so the eigenvectors can be mapped back to the Gaussian parameter space and the corresponding motion trajectories can be easily visualized. The first eigenvector from PCA shows a correlated motion of the helicase domain and the SF3b subunits, similar to the motion trajectories reported in previous studies.

While the eigenvectors from PCA exhibited several overall modes of motion of the complex, to better interpret the mechanism of the system, it is more interesting to look at the eigen-motion trajectories localized in individual domains. The use of PCA does not change the fact that the latent space has a non-linear relationship to the motions of the system. Thanks to the characteristics of Gaussian models, we can focus on specific regions in real space. Rather than decomposing the point cloud in the neural network latent space with PCA, we search for origin-crossing vectors in the latent space where the motion of Gaussian coordinates along the line is maximized in the domain of interest but minimized in rest parts of the protein (Fig 4). Since the points from this dataset form a relatively isotropic distribution in the latent space, the movement represented by these vectors are nearly as significant as the eigen-motion from PCA, while the Gaussian functions that are involved in the motion are much more localized. Furthermore, since the motion trajectories are localized in different domains, the two eigen-motion vectors are also orthogonal.

With the two independent Eigen-motion vectors localized in the helicase and SF3b domains, we investigated the coordination of the two domains by looking at motion trajectories produced by the linear combinations of the two vectors. Adding the two vectors results in a motion mode that the two domains are moving toward the same direction, similar to the first eigen-motion extracted by PCA from the system. In the alternative combination, the two domains can be seen move apart from each other, a motion mode never reported for the dataset (Fig 4f, Fig S4). Note that the individual presented structures are the 3-D reconstructions of particles near the corresponding point on the manifold. That is, unlike normal mode analysis, which predicts hypothetical modes with unknown amplitude and phase, in this case specific 3-D structures are generated from the data for each putative point, demonstrating that each specific state can be generated from the data, and that relative populations of different states can be considered, within the limits of noise.

**SARS-COV-2 spike protein**

Our third, somewhat timely, test system is the spike structure of SARS-Cov-2 (EMPIAR-10492). While the opening of the Receptor Binding Domains (RBD) was not observed in the deposited particles due to the sucrose cushion used in sample preparation[28], the RBDs in the final structure still have weaker density and lower resolution compared to the rest of the protein. 3-D classification was performed on the dataset in the original publication, but only an asymmetrical structure with weak RBD density was reported, and it was unclear in that study what conformational changes cause the weakening of the RBD density.

To investigate this question, we performed heterogeneity analysis on the combined particle set of the RBD closed and weaker density state (55,159 particles). To demonstrate that e2gmm is directly compatible with other software, rather than solving the structure again in EMAN2, we make use of the averaged structure and particle orientations from the deposited Relion refinement results. 2188 Gaussian functions were used to model the averaged structure at ~7 Å. To break the C3 symmetry, we treat every particle as three copies in the three symmetrical orientations, and only Gaussian functions in one of the asymmetrical units are allowed to vary from the neutral structure, so that every particle is mapped to three points in the conformation space, corresponding to the three asymmetric units.

After training, we performed PCA on the points in conformation space and the eigenvectors show the motion of secondary structure elements at the RBD. Along the first eigenvector, the alpha helix at residue 335-344 can be seen tilting toward the RBD of its adjacent subunit by ~11 degrees (Fig 5, Fig S8). Interestingly, in the averaged structures along the same trajectory, the same helix in one of the neighboring subunits is undergoing the same motion, but in the opposite direction (Fig 5f-h). Since the adjacent subunit was not targeted in the heterogeneity analysis, the presence of correlated motion suggests that the conformational changes of the RBDs in the two subunits are coordinated. Meanwhile, the same domain in the other subunit remains unchanged. On the other hand, the second eigenvector from PCA emphasizes the motion of the alpha helix at residue 364-371, as well as the beta-sheet strain at residue 354-359. Some coordination of motion in the adjacent subunit can also be observed but it is less clear.

In the density maps reconstructed from particles in specific conformations, the RBD at the subunit we focus on has stronger density than that of the other two subunits, suggesting the conformational changes we observe are indeed contributing to the weakening of density at the RBD (Fig S2).

**Discussion**

The major difference between the proposed method and the majority of CryoEM variability methods is the representation of the structure by a set of Gaussian functions, similar to methods sometimes employed in coarse-grained modeling[29,30]. This is analogous to the idea of directly refining the atomistic structure against the raw data, but in a reduced representation based on the scale of the expected variations. This representation provides a number of advantages. First, it greatly reduces the number of parameters that needed to represent the molecule at any specified level of detail, limited by the sampling of the image data[31]. For example, in the case of spliceosome, to represent the structure at 13Å using a voxel-based representation, the density map can be downsampled to a cube with a box size of 84, so a total of 592704 floating point parameters is required to represent the volume. Using the GMM, only 10240 variables are needed for the 2048 Gaussians used in the model, and the average FSC between the GMM and the density map is still above 0.95 for the spatial frequencies under consideration, indicating that it is a very good representation of the density map.

Second, at low resolution, Gaussian functions are a natural way to model CryoEM maps[32–34]. If these methods were extended to atomic resolution representation, it may be necessary to include the atomic form factors and consider the differences between electronic potential and electron density, but at intermediate resolutions these subtleties are effectively undetectable. Typical protein structural variability, such as ligand binding and domain motion, can be easily represented as simple trajectories in the Gaussian parameter space. Whereas under voxel-based representations, a high-dimensional nonlinear model is required to depict the motion of a domain along even a linear trajectory, especially when the length of the path is longer than the size of the domain of interest. As a result, the complexity of the model required to describe the structural variability of the protein is greatly reduced, easing the effort to train the encoder-decoder neural networks.

Third, due to the mathematical characteristics of Gaussian functions in both real and Fourier space, our representation avoids the artifacts produced by common image processing operations. For example, to focus the analysis on a specific domain, we can select the Gaussian functions of the domain, and only allow the parameters corresponding to those Gaussian functions to change during the training of the model. As all Gaussian functions still exist in the projection images, this will not introduce artifacts. The properties of Gaussian functions also ensure the model is always smooth at any snapshot during a continuous motion. Since the projection operation is performed by transforming the coordinates of the Gaussian functions, no interpolation artifacts are introduced by rotation or non-integral translation. This also makes it easier to apply constraints in both spaces when studying the structural variability of proteins, such as focusing on specific domains in real space or limiting to a range of Fourier frequencies.

Finally, the use of Gaussian models makes the output from the neural networks directly and intuitively interpretable, unlike the typical abstract spaces produced by other "manifold methods"[14,15]. Each point in the conformational space is mapped to a set of Gaussian parameters, which corresponds to a complete 3-D structure in one conformation. This means that for any given point, a representation can be generated either by reconstructing the particle image data in the vicinity of the point, or by directly converting the Gaussians into a density representation. The Gaussian map can provide a direct representation of the variations the network has learned, while the particle reconstructions can confirm that the actual 3-D intermediate structures exist and agree with the Gaussians. For any two selected points in the confirmation space, it is easy to visualize the differences between those points by plotting a trajectory of coordinate motion or amplitude change. This can be especially useful in identifying putative changes when there are insufficient particles in the conformation of interest to provide a true 3-D reconstruction at sufficient resolution. The Gaussian representation remains equivalently resolved at any point.

Some limitations remain in the current implementation of the method. First, since e2gmm requires the orientation of each particle as input, it only works in situations where a portion of the molecule is rigid enough that a reasonable neutral 3-D structure exists, and reasonable particle orientations can be determined. While this is a safe assumption for most SPR cases, it is also possible that the protein complex of interest is so heterogeneous that the alignment in the initial refinement fails entirely, and particle-projection mismatch is caused by misalignment instead of conformational differences. A potential solution to this problem is simultaneous training of particle orientation and GMM conformation. While this is possible, training the model to convergence is more challenging when this approach is used.

Second, the protocol normally begins with an averaged structure of all particles, assuming this represents the "neutral structure", which is then perturbed by the network. This assumption is not always true. When a domain motion is large enough, regions in the averaged map may be sufficiently spread in space that no Gaussian function is identified in that region when the neutral model is trained. As a result, the model excludes motion in that region since there are no Gaussians present to move. This can be corrected by selecting a better "neutral structure" with stronger density in this region. If dealing with a system with compositional variability, such as multiple ligands which may or may not be present, it is critical that the training volume be one with some density present for all ligands. This potential problem can also be reduced by building the neutral Gaussian model directly from aligned particles instead of the averaged structure, although this will incur a time penalty and may lead to a less robust neutral model.

Finally, GPU memory currently limits the size and resolution of the model. For example, a GPU with 11GB memory supports up to 3200 Gaussians with particles sampled at 128 x 128 pixels, and a batch size of 8 during training. This would be sufficient to represent the 50S ribosome at ~8 Å resolution, or smaller proteins at proportionally higher resolution. So, for many proteins, the method is currently limited to variations at the level of secondary structural features. This limitation is due to the Gaussian representation currently required by the underlying TensorFlow system. We expect that continuing evolution of GPU hardware as well as TensorFlow itself will remedy this problem in the near future without requiring other compromises in the method itself.

Despite these minor limitations, e2gmm represents an easy to use mechanism for exploring macromolecular variability in CryoEM with results which can be easily and intuitively interpreted. The user can define the resolution of interest, easily approaching features at any level of detail, within hardware limits. The next obvious development for e2gmm would be to operate on CryoET data, to permit similar studies in the context of the cellular environment, but technically this adaptation is not entirely straightforward to achieve due to the high noise levels in individual tilts and the increase in the amount of coordinated image data this would entail. All of the GMM operations are available through the program e2gmm_refine.py, and a graphical interface for interactive examination of results and exploring changes in parameters is provided by e2gmm.py. All of the necessary software is provided as part of EMAN2.91. A tutorial with sample data is available at http://eman2.org.

## Acknowledgement

This work was supported by NIH grant R01GM080139 to S.L. and computational resources were provided by BCM's CIBR center for Computational and Integrative Biomedical Research.

**Figures**

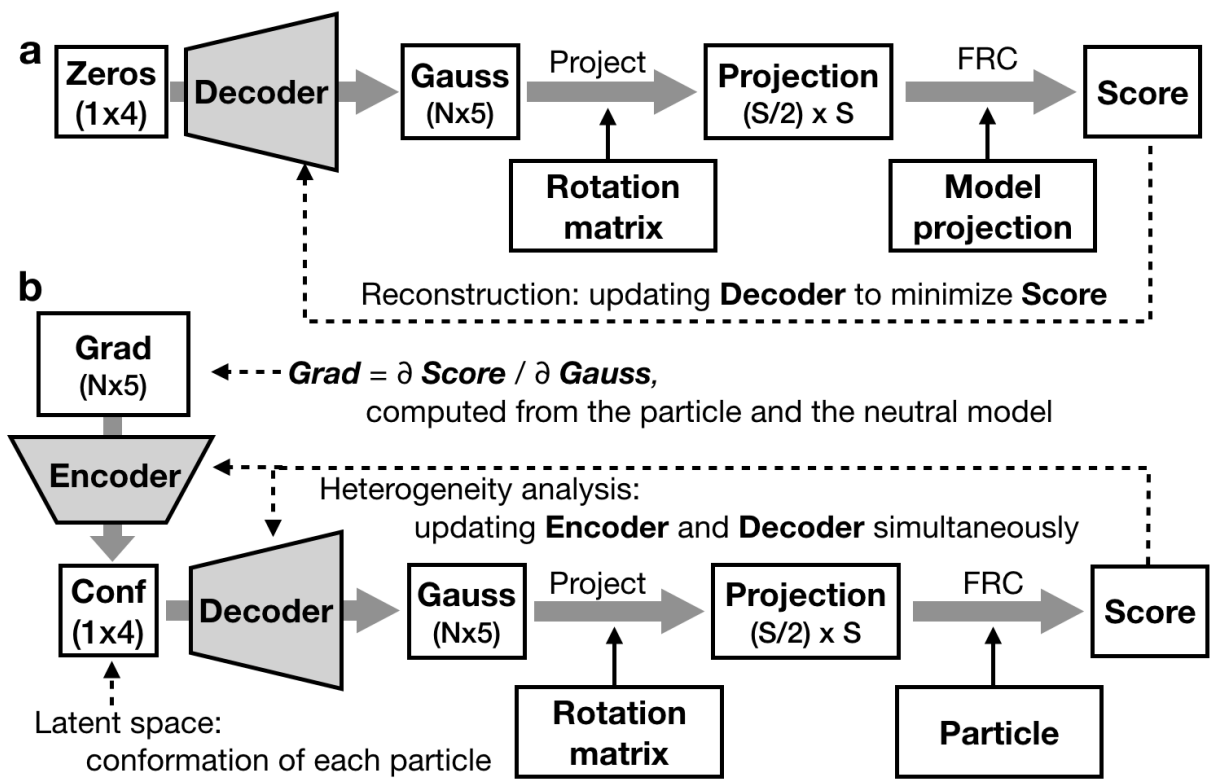

**Fig 1.** Neural network model. **(a)** Training the decoder to represent the neutral map. **(b)** Training the full network to represent system heterogeneity.

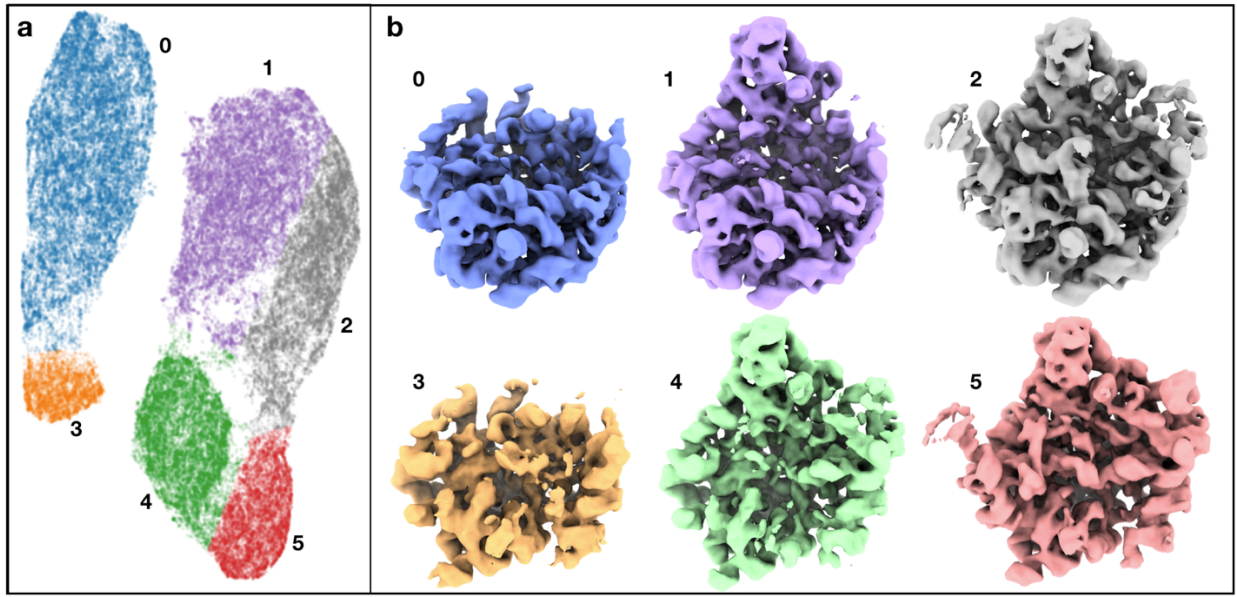

**Fig 2.** Classification of assembling ribosomes. **(a)** 2D embedding of particles from the 4-D latent space, colored by labels from clustering. **(b)** Averaged 3-D structures produced using the 2-D particles in each colored class, filtered to 8A. See also supplementary movie 1.

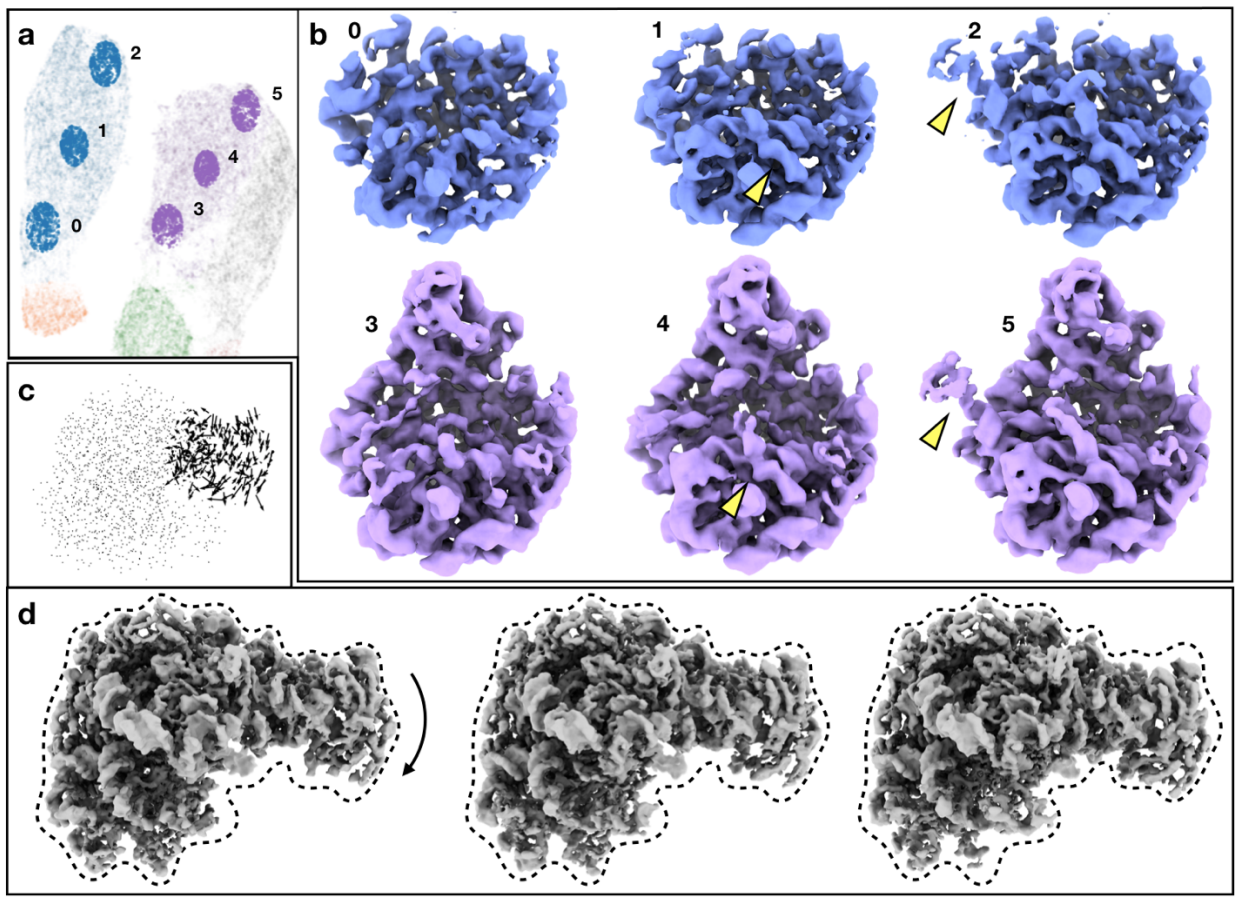

**Fig 3.** Exploration of subtle structural variability in the ribosome dataset. **(a)** Location of particles sampled from the 2D embedding of the conformation space. **(b)** Averaged structures reconstructed from the sampled particles. Yellow arrows point to the major differences between the structures. **(c)** Motion trajectories of Gaussian coordinates in the central protuberance domain from the first eigenvector of conformational heterogeneity analysis. **(d)** Averaged structures of the particles at points along the motion trajectory. The dotted envelope is fixed to better visualize the changes in each map. See also supplementary movie 2.

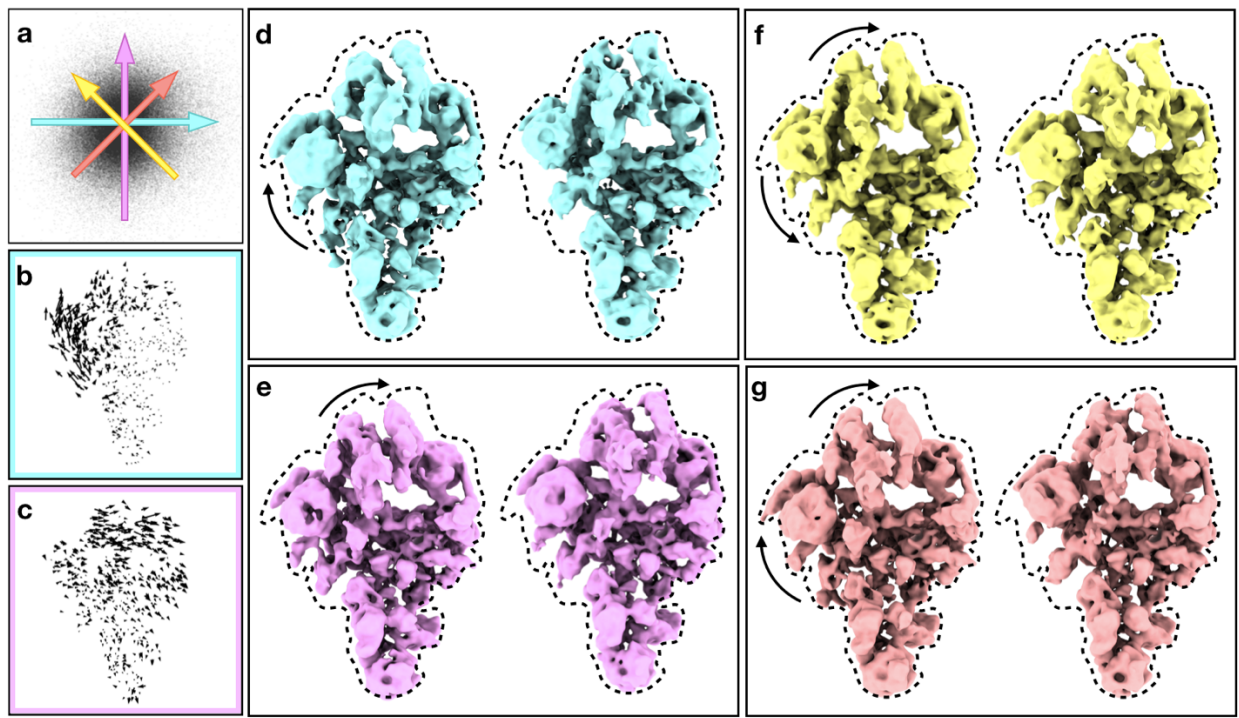

**Fig 4.** Structural variability analysis of spliceosome. **(a)** Distribution of particles in the 2D space formed by the selected base vectors. Colored arrows correspond to different motion trajectories shown in (d-g). **(b-c)** Motion trajectories of Gaussian coordinates from the two base vectors. Length of the vectors are exaggerated for better visualization. **(d-e)** Averaged structures reconstructed from particles along the two base vectors, showing the conformational change corresponding to **(b-c)** in 3-D. **(f-g)** Averaged structures reconstructed from particles along two different combinations of the base vectors, corresponding to the colored arrows in **(a)**. Dotted envelopes are fixed to better visually judge variations between maps. See also supplementary movie 3.

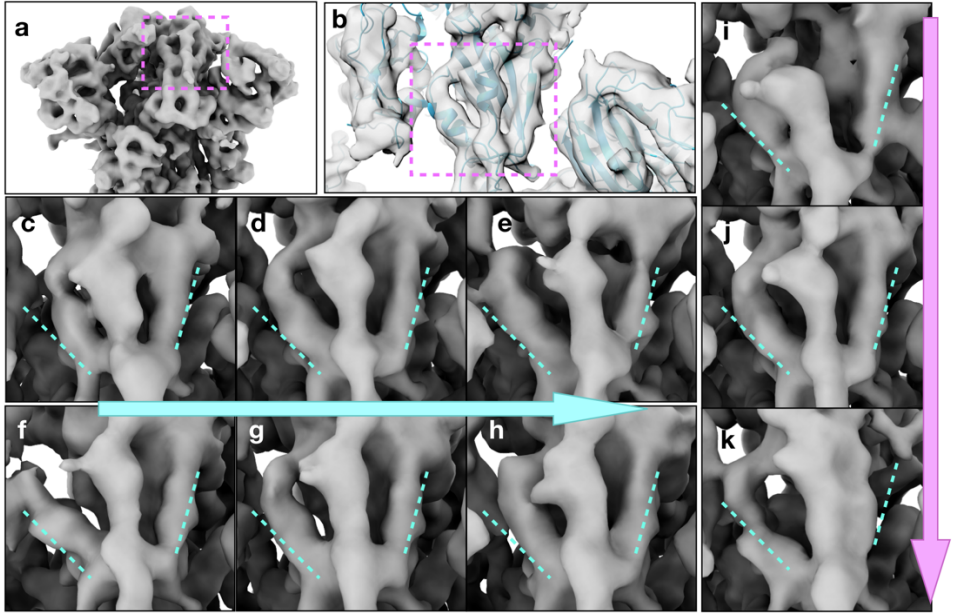

**Fig 5.** Structural variability analysis of the spike protein of SARS-COV-2. **(a)** Average structure of the spike protein, showing the RBD of the subunit that the analysis focuses on. **(b)** Density map of the target RBD overlaid with the molecular model (PDB: 6zwv). **(c-e)** Conformation change of the target RBD along the first eigenvector. **(f-h)** Conformation change of the one of the RBDs that are not targeted along the first eigenvector. **(i-k)** Conformation change of the target RBD along the second eigenvector. See also supplementary movie 4.

## Methods

**Gaussian representation of protein density maps**

The GMM is a simple sum of Gaussian functions in real space, $\bar{x} \in R^3$:

$$M(\bar{x}) = \sum_{j=1}^{N} A_j e^{-\frac{|\bar{x}-\bar{c}_j|^2}{2\sigma_j^2}}$$

Here, the Gaussian parameters are amplitude, $A_j$, width, $\sigma_j$, and center coordinates $\bar{c}_j$. In the network parameter space, the center parameters have a range (-0.5, 0.5), the amplitude has a range (0,1) and the Gaussian width (0.5, 1.5). Note that only the relative values of the amplitude and width of the Gaussian functions within the model are meaningful, as the FRC metric is insensitive to overall brightness and filtration of the imageThe center coordinates are scaled by the linear size of the image in pixels to form the projection images.

Internally a projection orientation is a 3x3 rotation matrix irrespective of the stored orientation in terms of Euler angles, quaternions, etc[35]. In the below equation, we discard the z component in the product, so R is 2x3, excluding the z row. A projection of the GMM in $\bar{t} \in R^2$ is thus simply:

$$P(\bar{t}) = \sum_{j=1}^{N} A_j e^{-\frac{|\bar{t}-R\bar{c}_j|^2}{2\sigma_j^2}}$$

Our loss function is the Fourier ring correlation (FRC) between the Fourier transform of $P$ and a particle image, $I$ in the same orientation. Note that the Fourier transform of $P$ can be computed by shifting the Gaussian sum to Fourier space for efficiency if the real-space representation is not required for some other purpose. The FRC between the Fourier transform of the GMM projection and a CryoEM particle image, is the average of the correlation coefficients over Fourier rings[36]:

$$\text{FRC}(\mathcal{P}, \mathcal{I}) = \frac{2}{b} \sum_{k=1}^{b/2} \frac{\sum_\theta \mathcal{P}_{k,\theta} \cdot \mathcal{I}_{k,\theta}}{\sqrt{\sum_\theta \mathcal{P}_{k,\theta}^2 \cdot \sum_\theta \mathcal{I}_{k,\theta}^2}}$$

where b is the box size in pixels, k,θ are FFT polar coordinates and $\mathcal{P}$ and $\mathcal{I}$ are the FFTs of $P$ and $I$. As this is an operation over the FFT of discrete images, the sum over θ covers all values at k±0.5 pixels. Since each ring is an independently normalized dot product, the FRC is insensitive to multiplication by any non-zero radial (filter) function. So long as the CTF phases have been correctly flipped and astigmatism and drift are minimal, CTF amplitude correction can be ignored. While the signal to noise ratio will be lower in the particle at points where the CTF amplitude is low, the FRC will still be maximized for an individual particle when the GMM best agrees with the underlying particle density irrespective of CTF.

**Neural network structure and parameter selection**

The structure of neural networks and the parameters during the two phases of training process are user adjustable, but the defaults are suitable for most use cases. By default, the encoder and decoder both have three fully-connected hidden layers, each with 512 units. A dropout layer with a rate of 0.3, as well as a batch normalization layer is included before the final output layer of both networks (Fig. S7). The ReLu function is used for activation in each layer, except for the output layer of the decoder, which uses a sigmoid activation function. During the training process, the default learning rate is 0.0001, with an L2 regularization of 0.001. A small random variable is also added to the latent space vector before it is input to the decoder, as a way to enforce the continuity of particle distribution in the latent space. This is similar to the concept of variational autoencoder, except that the variation of the random variable is not trainable here. An additional regularization factor is applied to the standard deviation of the amplitude and width of Gaussian functions to encourage the Gaussian functions to spread out in real space.

The number of Gaussian functions used in the model is decided based on the size of molecule and the target resolution. In practice, to build a Gaussian model from a density map, we start from a small amount of Gaussian functions (e.g.- 256) and target a low resolution (e.g.- 50 Å) and run the decoder optimization until the FRC curves between projections of the Gaussian model and the projections of the density map below the target resolution are always above 0.95. Then double the number of Gaussian functions, increase the target resolution and repeat the process. When increasing the number of Gaussian functions, each newly added Gaussian will be seeded near an existing one, so the low-resolution correlation between the Gaussian model and the density map from the previous round is roughly preserved. Typically, within 3-5 rounds, the decoder can produce a Gaussian model that matches the density map at the target resolution. When a user-defined mask is provided, the program will exclude Gaussian functions whose centers fall out of the mask, resulting in slightly fewer Gaussian functions than the targeted number.

Using the trained decoder, is possible to visualize any point in the latent space, or a derived reduced representation if it can be mapped back to the latent space. It is sometimes also useful to display the vector motions connecting two points in the latent space for all Gaussians. This can be easily presented as a quiver plot, with a vector drawn for each Gaussian between its position at point A in the latent space to its position at point B in the latent space. If the motions are particularly small compared to the size of the molecule, an optional scaling factor can be used to make the vectors more visible. A graphical tool, e2gmm_analysis.py, is provided to easily generate such plots.

**Tests on simulated datasets**

To verify the method's fundamental capabilities, testing was performed on three simple simulated datasets (Fig S5), each consisting of random projections of a dynamic 3-D model with a small amount of added noise. The first system included a large rigid domain with a smaller domain undergoing linear motion. The pathlength of the linear motion was longer than the width of the moving domain. Twenty 3D density maps were generated along the trajectory and 200 particles were generated for each 3D map, by projecting the map in a random orientation and adding a small amount of noise. In the simulation, we simply used the known projection orientations, since the routine is normally used with predetermined orientations. For simplicity, in this example we use a 1-dimensional latent space to avoid the need for any further dimensional reduction. After training the GMM, the resulting latent variable has good agreement with the location on the path. A plot of true conformation vs the single latent variable is shown in Fig S5d.

It is worth noting that even for this simple system, the estimated particle conformation distribution includes some off-diagonal points, which will tend to be biased towards zero, the neutral conformation. This is because the simulated domain movement occurs in a plane, so in some orientations the motion is effectively unobservable. In such cases there is a bias towards the neutral state. While this artifact is unavoidable and populating the manifold with particles will produce some near the origin of the latent space irrespective of their true conformation, this does not mean the manifold itself is inaccurate. So long as orientations are sufficiently diverse, the manifold should still be accurately determined. Indeed, with some effort it may be possible to remove such outliers from the particle distribution by testing whether the change in GMM would be detectable in the orientation of each individual particle.

In the second example, we simulated the cyclic rotation of a small domain around an axis, to show the method can learn nonlinear/cyclic motion trajectories. In the simulated dataset, 36 density maps were generated along the movement trajectory, and 200 particles were used for each snapshot. Here, we used a 2D latent space, so the motion could be directly modeled by the encoder with no further dimensional reduction. After training, the particles distribution in the latent space roughly formed a circle (Fig S5f), and when viewed in polar coordinates, the angle of each point in the latent space correlates well with its ground truth rotation angle of the small domain (Fig S5g).

Finally, we demonstrate the performance of the method when the system contains a mixture of conformational and compositional heterogeneity. The domain motion in this simulated system is the same as the first example, but for half of the population, we added a small additional density to the map, to represent compositional variability (Fig S5h). The compositional difference and the domain motion are independent. A Gaussian model was built from the averaged density map and trained to embed the particles onto a 2D latent space. After training, particles form two curves on the latent space that are roughly parallel to each other. Comparing to the ground truth conformation of the particles, it is clear that points on the two curves represent particles with and without the extra density, and the trajectory along the curve represents the linear motion of the flexible domain (Fig S5i). This also highlights the ability to separate compositional and conformational heterogeneity within the system.

**Additional data processing details for tests on real data**

For the ribosome dataset, obvious ice contamination was removed using the EMAN2 neural network particle picking tool prior to refinement. Single model refinement was performed using the remaining particles, which were split into two independent subsets. As we were not attempting to test the refinement pipeline, a high-resolution structure (EMD-8455) was lowpass filtered and phase randomized beyond 20Å to serve as an initial model for the refinement.

For the spliceosome dataset, all provided particles were used in the single model refinement. A high-resolution structure (EMD-3683) was lowpass filtered and phase randomized beyond 25Å to serve as the initial model for the refinement.

For the SARS-COV-2 spike protein dataset, Phenix real space refinement was performed to produce atomic models of the RBD for each frame of the continuous motion. Each density map was lowpass filtered to 5Å, and the RBD of the target asymmetrical unit was segmented in UCSF Chimera using the PDB model 6ZWV. Real space refinement was performed using the segmented RBD domains and the PDB model as the starting point.

**Reproducibility**

Since the method includes stochastic components, it is worth considering reproducibility. Towards this end, we tested the analysis of the EMPIAR 10076 data set by evenly separating the data into even/odd subsets. The entire processing pipeline, including single model refinement, the generation of Gaussian model and the heterogeneity analysis were performed independently on each subset. The GMM parameters and the training process for the two subsets were the same as described for the full dataset.

While the learned spaces are not identical due to the stochastic nature of the process, the number of clusters, the arrangement of the clusters on the manifold, and the number of particles within each cluster are equivalent (Fig S6). Further, after clustering the particles and reconstructing a 3D structure for each class, we can find a 1-1 match between the 3D class averages from the two subsets. The structures of the matched classes from particles different subsets are highly consistent, and FSC between the corresponding structures extend beyond 4Å.

This test establishes that functional reproducibility, while not guaranteed, is clearly possible in this method. We suggest that this even/odd split test, similar to the "gold standard" methods used for resolution testing in single model refinement, represents a reasonable test of the reproducibility of biological conclusions drawn from the results of the method.

**Computational requirements**

For EMPIAR-10076, starting from a completed single particle refinement, the first round of heterogeneity analysis, which focuses on only the amplitude changes of Gaussians required ~3 hours on a GeForce RTX 2080 TI GPU, including the training of the GMM model and the dimension reduction process. Less than 1 hour on a 12-core workstation was required to embed the encoder latent space in 2D, perform clustering, and reconstruct all of the 3-D density maps. The second round of heterogeneity analysis focusing on the conformational change in the central protuberance domain also required ~3 hours on the GPU and <1 hour on the 12-core workstation.

For the EMPIAR-10180 dataset, the heterogeneity analysis required ~3 GPU-h and ~30 CPU-h for the reconstruction of the density maps along the four reported motion trajectories.

The provided Relion alignment was used for the EMPIAR-10492 dataset. Heterogeneity analysis required ~2 GPU-h plus ~10 CPU-h.

**Comparison to existing methods**

The recently published heterogeneity methods, CryoDRGN[14], CryoSPARC 3DVA[15] and e2gmm (this paper), all made use of the ribosome (EMPIAR-10076) and spliceosome (EMPIAR-10180) as two of their examples, permitting users interested in comparing the methods to draw their own conclusions. The three methods use very different approaches to solve the same problem, and each has its own advantages, which we discuss briefly.

3DVA solves the structural variability of the protein complex using a linear subspace model. The nature of the method makes it difficult to represent large scale motions, where the trajectory of the conformational change is not linear with respect to the intensity of individual voxels. On the other hand, the linearity constraint also greatly simplifies the problem. So, when the heterogeneity within the system meets the linearity criteria, often the case in single particle analysis, it can produce accurate results very quickly. For example, to solve the motion within the spliceosome dataset, 3DVA takes ~3 GPU hours, similar to e2gmm processing time, but 20x faster than CryoDRGN. Its performance is best demonstrated in the ribosome assembly dataset, as the compositional variability within the complex is strictly linear with the intensity of the voxels. Comparing to our approach (Fig S3), the separation of classes is more obvious in their linear subspace, even without the extra UMAP embedding step. GMM clustering on the 3DVA latent space shows 7 ribosome classes, and 6 of them directly match the 6 classes from our result. The extra class identified by 3DVA is similar to the subtle changes from our result shown in Fig 2b, which do not form obvious clusters in the conformation space but can still be resolved in e2gmm with further analysis.

CryoDRGN uses an encoder-decoder deep neural network architecture conceptually similar to ours, but the underlying data representation uses a classical coordinate-based approach on the 3D density map. Compared to e2gmm, one advantage of CryoDRGN is its capability to generate neutral state structures from particles with pre-determined orientations, without the need for a reference 3D density map. This makes it possible to obtain distant states that are not covered in the averaged structure from the single particle refinement. For example, in the 50S ribosome assembly example, CryoDRGN is able to capture the small cluster of 70S ribosome (<1% of particles), an impurity of the dataset, in the embedded conformational space, which was not immediately detected in the results from 3DVA or our software.

In e2gmm, our use of a GMM representation has numerous advantages, including a reduction in time and resource requirements. In the same benchmark datasets, e2gmm is roughly as fast as 3DVA, and 10-20x faster than CryoDRGN, while producing qualitatively similar results. Also note that the tests of our software are performed on a consumer grade GPU (GTX2080), which only has ~1/3 memory and substantially lower performance than the hardware used in the CryoDRGN and 3DVA benchmarks.

One of the major difficulties in analyzing protein heterogeneity using other "manifold methods" is to interpret the particle distribution on the manifold and draw biological conclusions from the results. Generally, to interpret the structural difference between any two points on a manifold requires identifying particles near both points and reconstructing them in 3-D. With e2gmm, we can put any latent vector into the decoder and immediately have a set of Gaussian coordinates to display on the screen. The user can literally drag the mouse around the latent space and observe the changes in the underlying Gaussian model interactively. Further, with e2gmm a mask can be employed to define a subset of Gaussians to model during network training. Since the underlying Gaussian model is completely smooth, doing this doesn't introduce any edge artifacts into the system. While the capability of representing particles on the determined manifold is similar across all of these methods, with e2gmm it is much easier to find specific paths in a potentially multidimensional manifold which correspond to specific variations of interest. Thanks to the advantages, we are able to identify conformational changes from the two datasets that are not described in the previous work, such as the tilting of the central protrusion domain of the ribosome, and the independent movement of the helicase and SF3b domains in the spliceosome dataset.

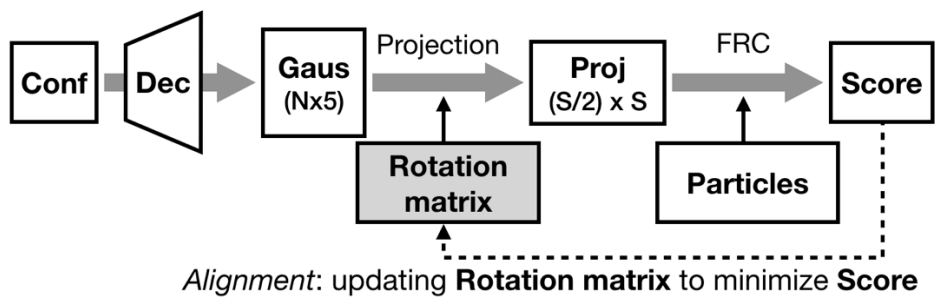

**Fig S1**. Workflow for local particle orientation refinement using the trained Gaussian model. This process can optionally be used after training the full GMM to improve particle orientations.

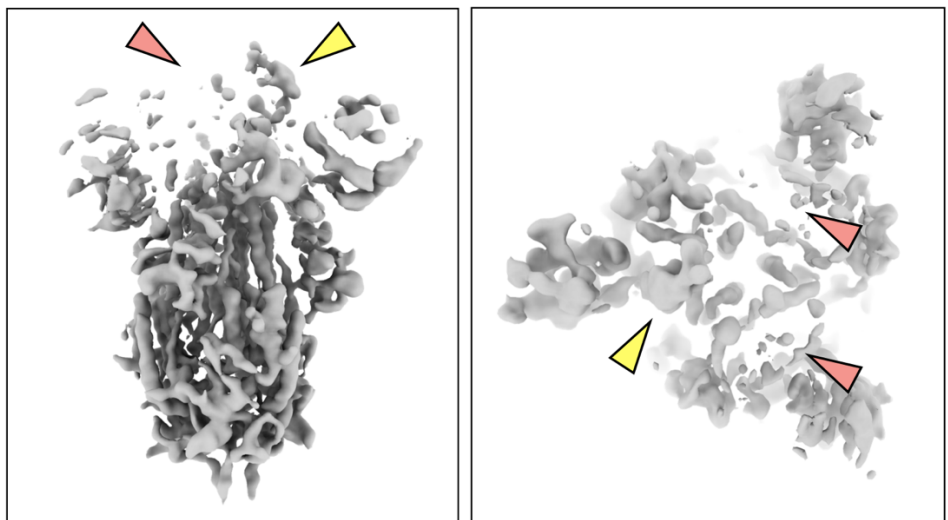

**Fig S2**. Structure of SARS-COV2 spike at one point in the continuous motion, visualized at high isosurface threshold. Note that the RBD of the subunit the heterogeneity analysis focuses on (yellow arrow) is still solid while the other two RBDs (red arrows) already vanish. This suggests the continuous motion is contributing to the weakening of density at the RBD.

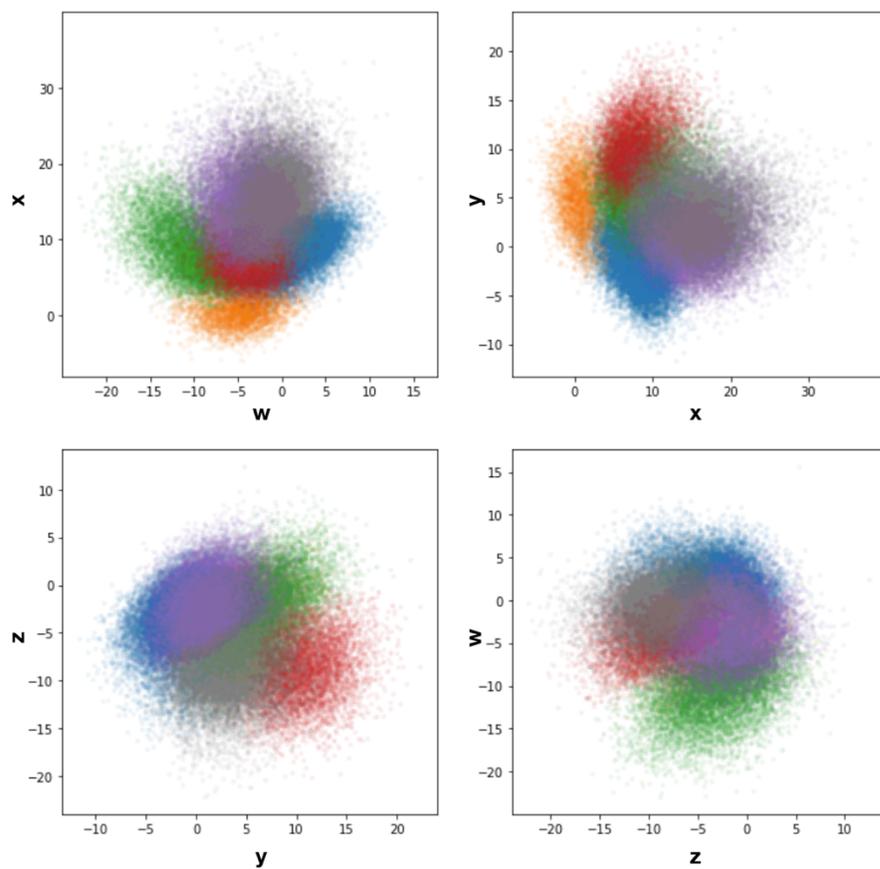

**Fig S3**. 50S ribosome particle distribution in the 4D encoder latent space, colored by the classification results shown in Fig 2.

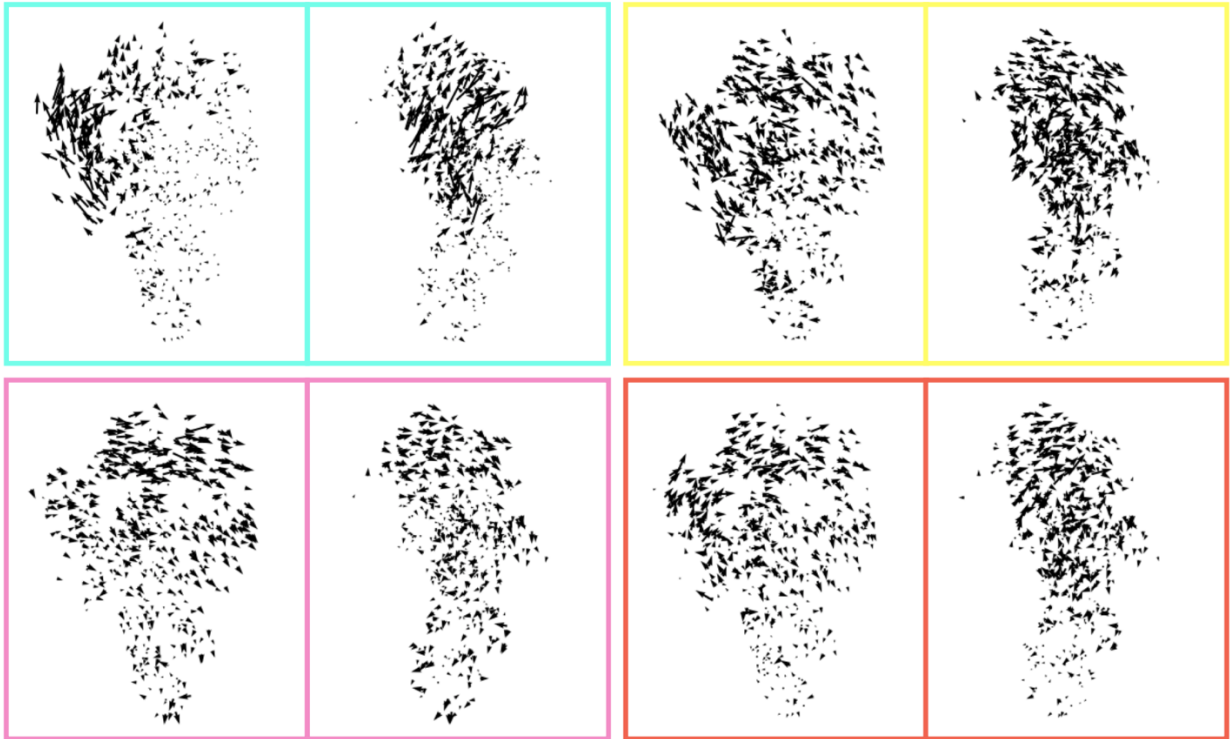

Fig S4. Front and side views of the motion trajectory vectors from the four identified motion modes of the spliceosome dataset shown in in Fig4 d-g.

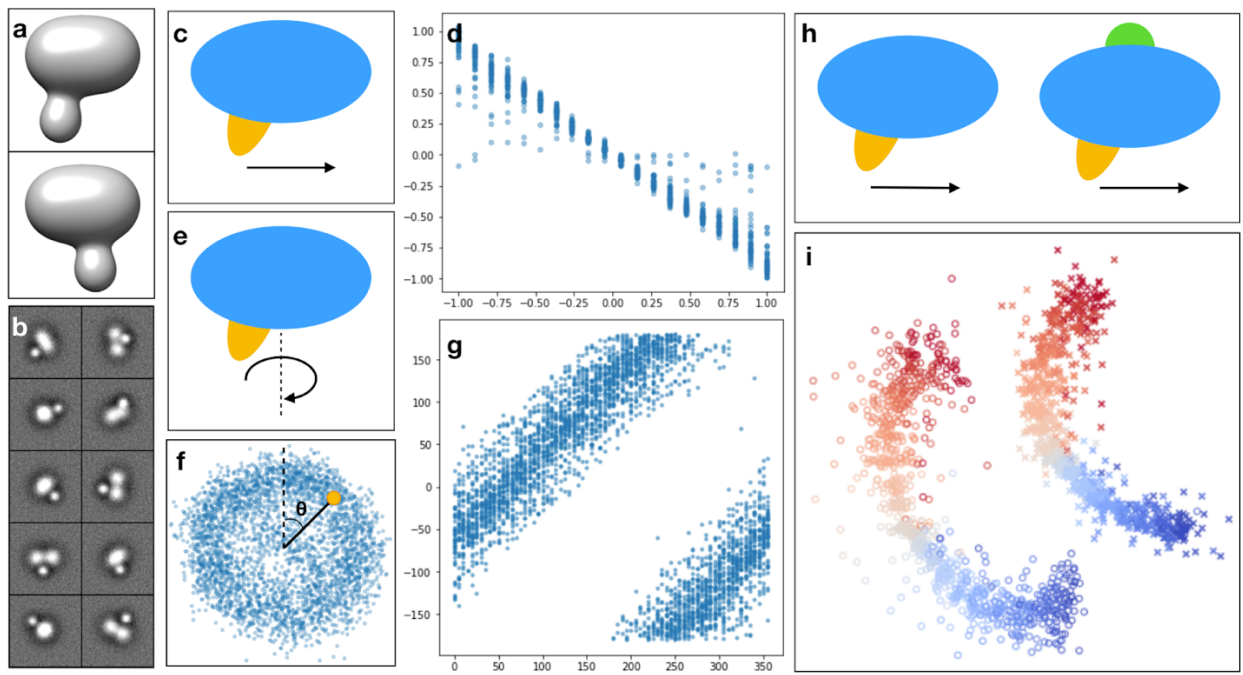

**Fig S5**. Results on simulated datasets. (a) 3D view of two snapshots of the simulated system at different frames of the movement trajectory. (b) Sample simulated particles. (c) Model of linear domain motion (yellow). (d) Scatter plot of the ground truth position vs estimated particle conformation of the linear domain movement. (e) Model of domain rotation around an axis. (f) Estimated particle distribution of (e) on the 2D latent space. (g) Scatter plot of the ground truth rotation angle vs estimated particle angle in the latent space (θ in f). (h) Combination of independent linear domain motion (yellow) and compositional change (green). (i) Particle distribution of (h) on the 2D latent space. Points are colored by their ground truth position along the linear domain motion trajectory. Particles with the extra density are marked as 'x', and the rest are marked as 'o'.

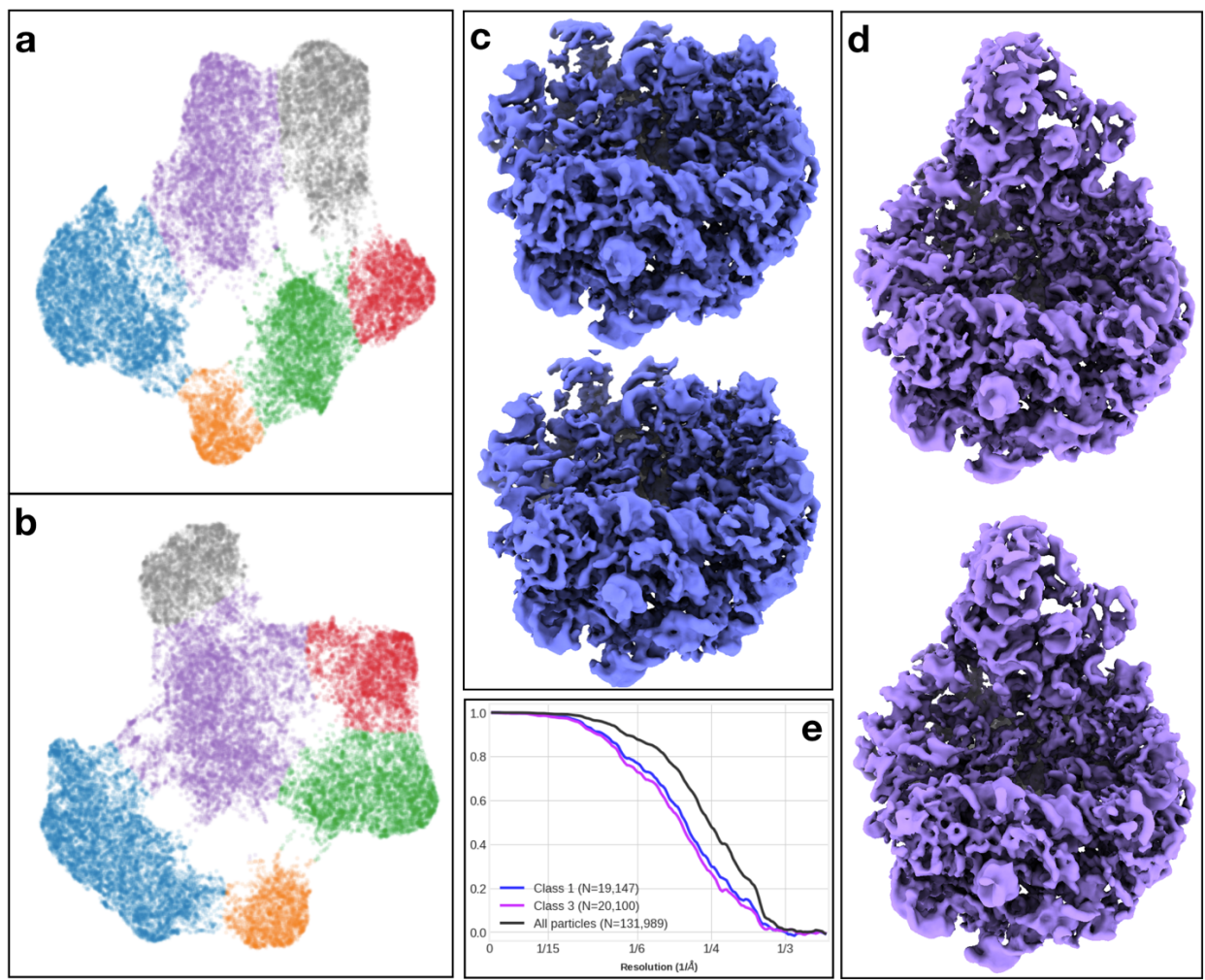

**Fig S6**. Reproducibility of the method on the ribosome dataset. (a-b) 2D conformation space embedding from heterogeneity analysis of two independent subsets of particles. The clusters are colored using the same scheme as Fig 2. (c-d) 3D class averages of particles in the same cluster from the two subsets. The maps are filtered by the local FSC between the two half maps. (e) "Gold-standard" FSC curves of the full dataset (black), and the two classes shown in (c-d) (blue and purple).

**Fig S7.** Detailed structure of the default neural network used for the examples shown in the paper.

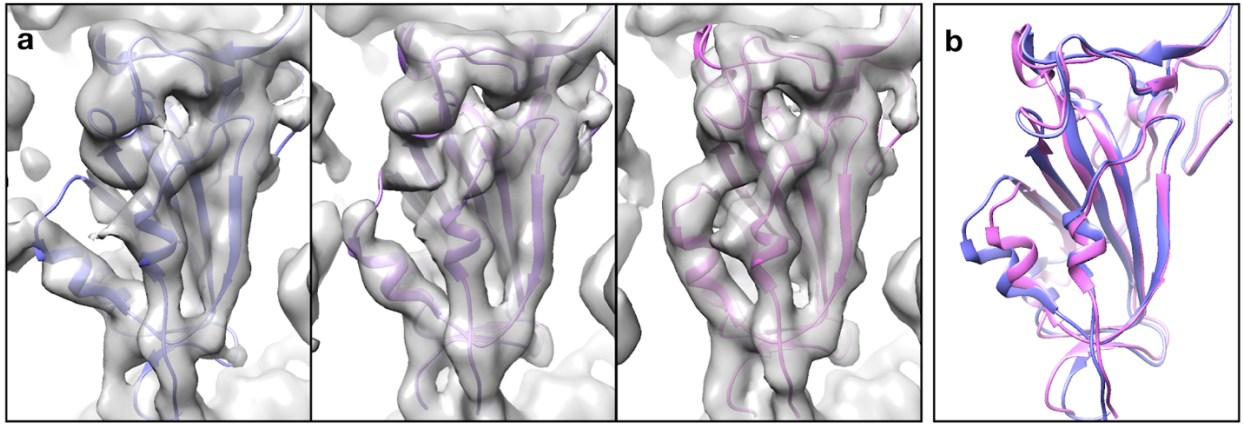

**Fig S8**. Molecular models fit to individual 3D snapshots of the focused RBD of the SARS-COV-2 spike protein, along the trajectory of the first eigenvector (Fig. 5c-e).